\documentclass[acmlarge,nonacm]{acmart}
\AtBeginDocument{\settopmatter{printacmref=true}}

\AtBeginDocument{%
  }

\setcopyright{acmlicensed}
\copyrightyear{2026}
\acmYear{2026}
\acmDOI{XXXXXXX.XXXXXXX}

\acmConference[XXX '26]{xxxx}{June 25 - 28}{Montreal, Canada}
\acmISBN{978-1-XXXX-XXXX-X/2026/01}

\usepackage{multirow}
\usepackage{enumitem}
\usepackage{svg}
\usepackage{subcaption}
\usepackage{fontawesome5}
\usepackage{etoc}
\usepackage{csquotes}
\usepackage{siunitx}
\DeclareSIUnit{\pp}{pp}
\begin{document}
\makeatletter
\addtocontents{toc}{\protect\etoc@depthtag{mainpaper}}
\makeatother

\title{The Collapse of Heterogeneity in Silicon Philosophers}

\author{Yuanming Shi}
\email{jeremyshi@adobe.com}
\affiliation{%
  \institution{Adobe Inc.}
  \city{San Jose}
  \state{California}
  \country{USA}
}

\author{Andreas Haupt}
\email{h4upt@stanford.edu}
\affiliation{%
  \institution{Stanford University}
  \department{Departments of Economics and Computer Science}
  \city{Stanford}
  \state{California}
  \country{USA}
}

\renewcommand{\shortauthors}{Shi, Haupt}

\begin{abstract}
  Silicon samples are increasingly used as a low-cost substitute for human panels and have been shown to reproduce aggregate human opinion with high fidelity. We show that, in the alignment-relevant domain of philosophy, silicon samples systematically collapse heterogeneity. Using data from $N = \num{277}$ professional philosophers drawn from PhilPeople profiles, we evaluate seven proprietary and open-source large language models on their ability to replicate individual philosophical positions and to preserve cross-question correlation structures across philosophical domains. We find that language models substantially over-correlate philosophical judgments, producing artificial consensus across domains. This collapse is associated in part with specialist effects, whereby models implicitly assume that domain specialists hold highly similar philosophical views. We assess the robustness of these findings by studying the impact of DPO fine-tuning and by validating results against the full PhilPapers 2020 Survey ($N = \num{1785}$). We conclude by discussing implications for alignment, evaluation, and the use of silicon samples as substitutes for human judgment. The code of this project can be found at ~\href{https://github.com/stanford-del/silicon-philosophers}{\faGithub}.
\end{abstract}

\ccsdesc[500]{Computing methodologies~Natural language processing}
\ccsdesc[300]{Computing methodologies~Philosophical/theoretical foundations of artificial intelligence}
\ccsdesc[300]{Computing methodologies~Knowledge representation and reasoning}

\keywords{large language models, silicon sampling, algorithmic fidelity, expert simulation, philosophy}

\maketitle

\section{Introduction}

Silicon sampling---conditioning large language models (LLMs) on demographic and attitudinal information to generate synthetic human responses---offers a low-cost alternative to human panels in social science research. \citet{argyle2023out} demonstrated that GPT-3 exhibits algorithmic fidelity when simulating U.S. political opinions, replicating not merely individual preferences but complex correlation patterns between attitudes. \citet{park2024generative} extended these findings through interview-based conditioning, achieving accuracy comparable to human test-retest reliability.

However, recent work raises concerns about whether silicon samples faithfully preserve heterogeneity within human populations. \citet{santurkar2023whose} found that LLM opinions systematically diverge from the general U.S. population. \citet{durmus2023towards} showed that models struggle to faithfully represent minority viewpoints. \citet{wang2025llm} demonstrated that LLMs flatten identity groups, amplifying stereotypes rather than capturing authentic diversity. These findings suggest that silicon sampling may produce artificial consensus where genuine disagreement exists.

Professional philosophy provides an ideal test case for whether silicon sampling preserves heterogeneity in expert-led but disagreement-rich communities. The PhilPapers Survey \citep{bourget2023philosophers} documents \num{1785} philosophers' positions across metaphysics, epistemology, ethics, philosophy of mind, and other subdomains. The survey reveals structured disagreement: philosophical positions exhibit strong correlation structures (e.g., physicalists tend toward atheism and naturalism), and the level of agreement varies substantially across questions.

This paper investigates whether LLMs can preserve these patterns when simulating philosophers, including whether they maintain the natural heterogeneity observed in human populations and whether they replicate the correlation structure of philosophical positions. We find that LLMs systematically collapse heterogeneity: they over-correlate philosophical judgments, producing artificial consensus, and exhibit spurious specialist effects whereby models assume domain specialists hold stereotypically aligned views. We investigate two complementary dimensions of representational fidelity:

\paragraph{RQ1 (Heterogeneity and Structural Alignment):} Do silicon samples exhibit the same magnitude and structure of philosophical disagreement as human philosophers?

\paragraph{RQ2 (Correlation Structure Preservation):} Do silicon samples preserve the empirically observed correlation structures between philosophical positions, and can fine-tuning improve this preservation?

To answer RQ1, we examine overall variance levels, domain-level predictability, demographic-position correlations, and whether the latent dimensions (that is, principal components) of disagreement correspond between LLMs and humans. For RQ2, we test whether LLMs maintain internal coherence across questions. Human philosophers exhibit significant correlations between related positions, and models that predict individual answers correctly but fail to preserve these correlations may indicate they lack genuine philosophical reasoning. Together, these questions carry implications for AI alignment: if LLMs collapse the heterogeneity of expert philosophers, who are arguably best equipped to provide judgments on alignment topics, they risk imposing artificial consensus where genuine disagreement exists. Our findings show systematic failures on both dimensions: LLMs produce $1.9$ to $3.9$ times lower variance than human philosophers and organize disagreement along fundamentally different axes. While fine-tuning with Direct Preference Optimization (DPO) improves correlation structure preservation, it does not resolve the underlying heterogeneity collapse.

This paper includes the following contributions: (1) a systematic test of algorithmic fidelity in an expert domain; (2) a new evaluation framework for heterogeneity preservation using principal component analysis following \citet{bourget2023philosophers}; (3) identification of systematic over-correlation and spurious specialist effects across all tested models; and (4) analysis of whether finetuning can remedy these concerns.

\begin{figure*}[t]
\centering
\includegraphics[width=\textwidth]{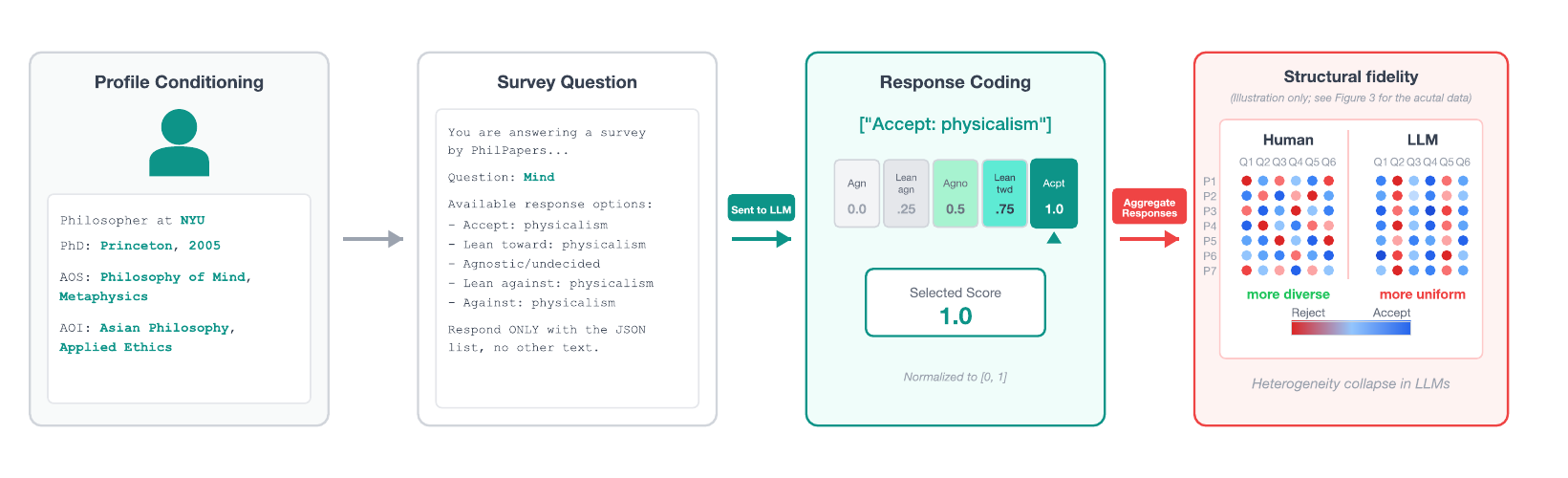}
\caption{Silicon sampling workflow: profile conditioning, survey question, response coding, and structural fidelity analysis.}

\Description{A four-stage diagram showing the silicon sampling workflow: profile conditioning with philosopher demographics, survey question presentation, response coding to normalized scores, and structural fidelity analysis comparing LLM versus human responses.}
\label{fig:overview}
\end{figure*}

\section{Related Work}

\paragraph{Silicon Sampling.}
\citet{argyle2023out} operationalize algorithmic fidelity through four criteria: (1) \emph{Social Science Turing Test}---generated responses are indistinguishable from human texts; (2) \emph{Backward Continuity}---responses are consistent with the socio-demographic conditioning context; (3) \emph{Forward Continuity}---responses proceed naturally from the context, reflecting appropriate form and tone; and (4) \emph{Pattern Correspondence}---responses reflect underlying patterns of relationships between ideas, demographics, and behavior observed in human data. Pattern correspondence is most relevant here, as it tests whether models capture structured belief systems rather than isolated responses. Their work on U.S. voter simulation established that LLMs can replicate complex correlations between political attitudes, demographics, and behaviors. \citet{park2024generative} extended this through interview-based conditioning with \num{1052} individuals, achieving \qty{85}{\%} accuracy, comparable to human test-retest reliability. Beyond political opinions, \citet{aher2023using} introduced ``Turing Experiments,'' demonstrating that models reproduce known human behavioral patterns in ultimatum games and prisoner's dilemmas. \citet{horton2024homo} proposed \emph{homo silicus}---LLM-based simulated economic agents, demonstrating their utility for pilot experiments in economics. \citet{sarstedt2024silicon} provide guidelines for silicon sampling in consumer research, noting that results vary considerably across domains. However, these investigations examined general populations on commonplace topics accessible to non-experts.

\paragraph{Persona Prompting.}
While silicon sampling shows promise for aggregate simulation, concerns about within-group heterogeneity have emerged. \citet{santurkar2023whose} documented systematic liberal leanings in LLM opinions. \citet{durmus2023towards} showed that opinion distributions vary substantially across models and can be shifted through prompting, but minority viewpoints remain underrepresented. \citet{wang2025llm} provided direct evidence of heterogeneity collapse, finding that LLMs reduce within-group variation and amplify stereotypes rather than capturing authentic diversity. \citet{sourati2026homogenizing} argue that LLMs are homogenizing human expression and thought at scale, reshaping linguistic norms beyond laboratory settings. \citet{bisbee2024synthetic} confirmed these concerns in the survey domain, showing that LLM-generated responses systematically distort opinion distributions. \citet{gao2024caution} further documented that LLMs overestimate effect sizes in experimental manipulations by a factor of two to ten, cautioning against their use as human surrogates.

Recent work has explored fine-tuning to mitigate these shortcomings. \citet{suh2025language} fine-tune on \num{3362} survey questions with response distributions from \num{70000} demographic profiles, improving distributional alignment \citep{sorensen2024pluralistic} . \citet{hewitt2024predicting} use LLMs to predict social science experiment outcomes. \citet{kolluri2025finetuning} fine-tune Llama-3.1-8B and Qwen-2.5-14B with both supervised fine-tuning (SFT) and Direct Preference Optimization (DPO) \citep{rafailov2023direct} on 2.9 million social science responses, finding that DPO yields the best individual-level prediction accuracy with robust generalization to unseen studies. However, \citet{kirk2024understanding} show that Reinforcement Learning from Human Feedback (RLHF) significantly reduces output diversity compared to SFT, implying a fundamental trade-off between alignment and diversity. Our work differs by focusing on an expert domain (philosophy) where structured disagreement instead of consensus is the phenomenon of interest.

\paragraph{Philosophical Disagreement.}
The PhilPapers Surveys \citep{bourget2014philosophers, bourget2023philosophers} document \num{1785} professional philosophers' positions on 100 questions, providing uniquely rich data for studying expert positions. Specifically, the survey reveals correlations between philosophical views, specialist effects and structured disagreement that does not organize along a single ideological axis, identified by principal component analysis (PCA).  \citet{schwitzgebel2012expertise} documented that professional philosophers exhibit cognitive biases despite expertise, suggesting philosophical reasoning combines systematic frameworks with persistent heuristics. \citet{cheung2025amplified} demonstrate that LLMs exhibit amplified cognitive biases in moral decision-making---with omission biases stronger than in humans---suggesting that normatively-loaded domains may be particularly susceptible to distortion. While these studies examine LLM biases in philosophical reasoning, none systematically test whether silicon sampling preserves the structured disagreement manifested among philosophers.

\section{Methods}

\subsection{Data}

The PhilPapers 2020 Survey \citep{bourget2023philosophers} reports aggregate distributions but does not release individual-level responses linked to demographic profiles. Silicon sampling requires individual conditioning, so we construct our own ground truth by compiling 277 publicly available philosopher profiles from PhilPeople \citep{philpeople}, yielding \num{5870} survey responses across 100 questions. Each profile includes areas of specialization (AOS), areas of interest (AOI), PhD institution/country, graduation year, and self-reported positions (Accept, Lean toward, Agnostic, Lean against, Reject). Our sample is drawn from 129 PhD institutions across 16 countries with a North American skew that we discuss as a limitation; topical diversity tracks the PhilPapers 2020 Survey: across the ten most common AOS categories in \citet{bourget2023philosophers}, our sample shares differ by 3.4 percentage points on average (max 9.2 pp for Social and Political Philosophy). For fine-tuning (see Section~\ref{sec:finetuning} for preference-pair construction), we use the PhilPapers 2009 Survey results \citep{bourget2014philosophers}, which preserve correlation structures among different philosophical positions despite lacking explicit individual demographic data.

For matrix similarity (Section~\ref{sec:matrix_sim}), PCA (Section ~\ref{sec:PCA}), and question correlation structure (Section~\ref{sec:question_corr}), we apply a normalization method largely following \citet{bourget2023philosophers}: one variable per question (the most popular option among human respondents for non-binary questions; the positive option with complement recovery for binary questions), yielding a $277 \times 100$ matrix per data source. 

Table~\ref{tab:model_summary} summarizes the response statistics across all evaluated models using this normalization. Human data exhibits high missingness (\qty{61.1}{\%} after complement recovery) as philosophers selectively answer questions in their areas of expertise; LLM data shows lower missingness (\qty{17}{\%}--\qty{40}{\%}), reflecting parsing failures and refusals.

\begin{table}[t]
\begin{minipage}[c]{0.52\columnwidth}
\centering
\footnotesize
\setlength{\tabcolsep}{3pt}
\begin{tabular}{@{}lccccc@{}}
\toprule
Model & N & Q & Responses & Resp\% & Per-Q Var \\
\midrule
Human & 277 & 100 & \num{10768} & \qty{38.9}{\%} & 0.053 \\
\midrule
GPT-4o & 277 & 100 & \num{16670} & \qty{60.2}{\%} & 0.014 \\
GPT-5.1 & 276$^\dagger$  & 100 & \num{16526} & \qty{59.9}{\%} & 0.014 \\
Claude-Sonnet-4.5 & 277 & 100 & \num{16612} & \qty{60.0}{\%} & 0.026 \\
Llama-3.1-8B & 277 & 100 & \num{19144} & \qty{69.1}{\%} & 0.026 \\
Llama-3.1-8B (FT) & 277 & 100 & \num{18413} & \qty{66.5}{\%} & 0.019 \\
Mistral-7B & 277 & 100 & \num{18764} & \qty{67.7}{\%} & 0.028 \\
Qwen-3-4B & 277 & 100 & \num{23077} & \qty{83.3}{\%} & 0.016 \\
\bottomrule
\end{tabular}
\end{minipage}%
\hfill
\begin{minipage}[c]{0.44\columnwidth}
\caption{Model response statistics. $N$ = philosophers, $Q$ = questions, Resp\% = response rate, Per-Q Var = average within-question variance. Human counts reflect effective coverage after complement recovery on binary stems. $^\dagger$One silicon philosopher returned empty results.}
\label{tab:model_summary}
\end{minipage}
\end{table}

\subsection{Evaluation}
\label{sec:evaluation}

We employ complementary metrics from ecology, information theory, and multivariate statistics, each capturing different aspects of fidelity.


\paragraph{Specialist Effects.} We validate LLM specialist effects against two sources: PhilPapers 2020 Survey ($N=\num{1785}$) and our 277-philosopher ground truth. Two-source validation provides stronger evidence for spurious associations, avoiding conflation of sample-specific artifacts with genuine LLM biases. For each specialist effect, we compute chi-squared statistics with Yates correction. An effect is \emph{spurious} if the ground truth shows no significance ($p > 0.05$) but LLMs show significant associations.

\paragraph{Matrix Correlation.} To assess whether LLMs preserve relationships between philosophical questions, we compute question-to-question correlation matrices using pairwise deletion (each pair of questions uses only philosophers who answered both), then compare these matrices. Philosophical coherence manifests in systematic relationships between positions; models that predict individual answers correctly but fail to preserve correlation structures would lack genuine philosophical reasoning. We use the Mantel test \citep{mantel1967detection}, a permutation-based procedure that assesses whether two correlation matrices are more similar than expected by chance (significance established via 999 random permutations), and the RV coefficient \citep{robert1976unifying}, defined as $\text{RV}(X, Y) = \mathrm{trace}(XX'YY') / \sqrt{\mathrm{trace}(XX')^2 \cdot \mathrm{trace}(YY')^2}$, a multivariate generalization of $R^2$ ranging from 0 (no similarity) to 1 (identical structure).

\paragraph{Information-Theoretic Measures.} We employ Kullback-Leibler (KL) divergence \citep{kullback1951information} and Jensen-Shannon (JS) divergence \citep{lin1991divergence} in two contexts: (1) comparing distributions of pairwise correlation coefficients between question-to-question correlation matrices (Section~\ref{sec:question_corr}), and (2) comparing flattened response matrices element-wise (Section~\ref{sec:matrix_sim}). For correlation-distribution comparisons, we discretize correlation values into 20 bins spanning $[-1, 1]$; for response-matrix comparisons, we discretize response values into 20 bins spanning $[0, 1]$. In both cases we compute:
\begin{equation*}
D_{\text{KL}}(P \| Q) = \sum_i P(i) \log \frac{P(i)}{Q(i)}
\end{equation*}
\begin{equation*}
D_{\text{JS}}(P \| Q) = \frac{1}{2}D_{\text{KL}}(P \| M) + \frac{1}{2}D_{\text{KL}}(Q \| M)
\end{equation*}
where $M = \frac{1}{2}(P + Q)$. Lower values indicate closer distributional match to human data.

\paragraph{Response Diversity.} We measure response diversity using Shannon entropy \citep{shannon1948mathematical} of the response distribution per question. Heterogeneity collapse manifests as reduced entropy: models generating near-uniform responses where humans exhibit genuine disagreement. We compute:
\begin{equation*}
H(X) = -\sum_{i=1}^{n} p(x_i) \log_2 p(x_i)
\end{equation*}
where $p(x_i)$ is the proportion of responses taking value $x_i$. Higher entropy indicates more diverse responses; entropy of 0 indicates uniform responses. We track the percentage of questions with non-zero variance.

\paragraph{Principal Component Analysis.} To characterize the dimensionality and structure of philosophical disagreement, we apply principal component analysis following the methodology of \citet{bourget2023philosophers}: for each question, we select one numerical variable (choosing the most popular option among human respondents for non-binary questions), yielding 100 variables per data source.

PCA is appropriate because philosophical positions are not independent. Beliefs form coherent worldviews with systematic correlation patterns, and PCA reveals whether LLMs preserve this structured disagreement or collapse it into artificial consensus. Following \citet{bourget2023philosophers}, who used R's \texttt{missMDA} package for imputation, we apply iterative PCA imputation: initialize missing values with column means, fit PCA with 5 components, reconstruct and replace only missing entries, and iterate until convergence. This avoids the information loss of listwise deletion while producing a complete data matrix suitable for standard PCA. Moreover, data imputation is used only for PCA; question correlation structure analysis (Section~\ref{sec:question_corr}) uses pairwise-deletion correlation matrices to avoid introducing artificial structure.

Human responses have substantial missingness (\qty{61.1}{\%} after complement recovery), while LLM data has lower missingness (\qty{17}{\%}--\qty{40}{\%} depending on model). Iterative PCA imputation handles this asymmetry by leveraging the correlation structure within each dataset to fill missing values, rather than fabricating independent positions.

Following \citet{bourget2023philosophers}, we count components that explain at least $\qty{2}{\%}$ of variance each. While PCA mathematically extracts as many components as there are variables, only components exceeding this threshold represent meaningful dimensions of disagreement rather than noise. We report: (1) variance explained by each component, (2) number of components exceeding the \qty{2}{\%} threshold, (3) top-loading questions for each component, and (4) loading correlations between human and LLM components after optimal alignment.

\subsection{Language Models}

We evaluate seven language models: proprietary (GPT-5.1, GPT-4o, Claude-Sonnet-4.5), open-source (Llama-3.1-8B, Qwen-3-4B, Mistral-7B), and a DPO fine-tuned variant of Llama-3.1-8B.\footnote{Exact snapshots: \texttt{gpt-4o-2024-11-20}, \texttt{gpt-5.1-2025-11-13}, \texttt{claude-sonnet-4-5-20250929}, \texttt{meta-llama/Llama-3.1-8B-Instruct}, \texttt{mistralai/Mistral-7B-Instruct-v0.3}, \texttt{Qwen/Qwen3-4B}.} Following \citet{argyle2023out}, we condition each model on philosopher profiles to generate synthetic survey responses using persona and question prompts (Appendix~\ref{app:prompts}); a prompt sensitivity analysis confirms our findings are robust to alternative framings (Appendix~\ref{app:prompt_sensitivity}). We use temperature 0 to isolate model knowledge from sampling stochasticity. Responses are coded on an ordinal scale from $+2$ (Accept) to $-2$ (Reject), normalized to $[0,1]$. Closed-source models achieved \qty{100}{\%} parsing success; open-source models ranged from \qty{89.4}{\%} to \qty{99.9}{\%} (Appendix~\ref{app:parsing}).

\subsection{Fine-tuning}
\label{sec:finetuning}

To investigate whether fine-tuning can improve structural fidelity, we apply DPO \citep{rafailov2023direct} to Llama-3.1-8B. DPO is particularly suitable for this domain because philosophical positions are matters of informed preference rather than objective truth---philosophers endorse positions based on reasoned judgment, not factual correctness. Unlike SFT, which treats philosophical positions as ground truth labels, DPO's preference-learning framework better captures the nature of philosophical disagreement. We empirically validate this choice: SFT on the same data produces severe mode collapse (\qty{84}{\%} of responses default to ``Accept''; see Appendix~\ref{app:sft_vs_dpo}). 

Mathematically, DPO optimizes the policy to prefer chosen over rejected responses via a contrastive objective:
\begin{equation*}
\mathcal{L}_{\text{DPO}}(\pi_\theta; \pi_{\text{ref}}) = -\mathbb{E}_{(x,y_w,y_l)}\Bigl[\log \sigma\Bigl(\beta \log \frac{\pi_\theta(y_w|x)}{\pi_{\text{ref}}(y_w|x)} - \beta \log \frac{\pi_\theta(y_l|x)}{\pi_{\text{ref}}(y_l|x)}\Bigr)\Bigr]
\end{equation*}
where $x$ is the context (in our case sociodemographics and the question), $y_w$ and $y_l$ are chosen and rejected responses (in our case tokens corresponding to answers), $\sigma$ is the sigmoid function, $\pi_{\text{ref}}$ is a reference model, in this case Llama-3.1-8B, and $\pi_\theta$ is the model whose weights are finetuned.

We construct preference pairs from philosopher survey responses: for each philosopher's actual position (chosen), we generate a rejected response by inverting the stance (e.g., Accept $\leftrightarrow$ Reject, Lean Accept $\leftrightarrow$ Lean Reject). Agnostic responses are skipped because they have no natural inversion. Training uses \num{3434} examples from 226 philosophers drawn from the PhilPapers 2009 Survey \citep{bourget2014philosophers}---a separate sample from our 277 evaluation philosophers, who are drawn from the 2020 survey. Of the 226 training philosophers, some overlap with our evaluation set, but the two samples were collected independently---one in 2009, and the other in 2020. We apply Low-Rank Adaptation (LoRA) \citep{hu2021lora} for parameter-efficient fine-tuning with rank $r=16$, $\alpha=32$, and dropout 0.05. Training runs for 2 epochs with batch size 1, gradient accumulation over 4 steps, learning rate \num{5e-6}, and DPO regularization $\beta=0.1$.

\section{Results and Discussion}

\begin{figure}[t]
\centering
\begin{subfigure}[b]{0.75\columnwidth}
\includegraphics[width=\textwidth]{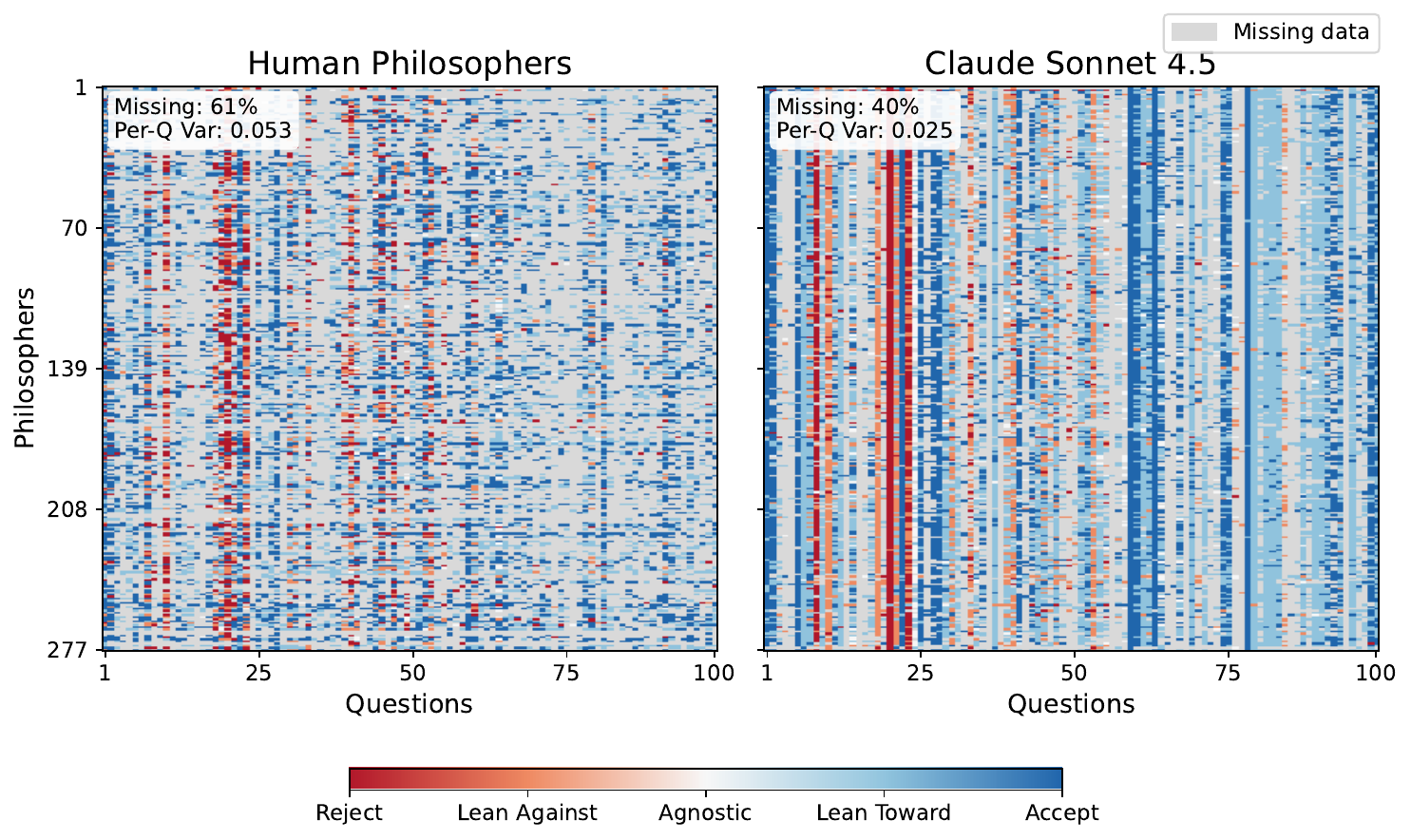}
\caption{Human vs.\ Claude-Sonnet-4.5}
\label{fig:human_sonnet}
\end{subfigure}

\vspace{0.3em}

\begin{subfigure}[b]{0.75\columnwidth}
\includegraphics[width=\textwidth]{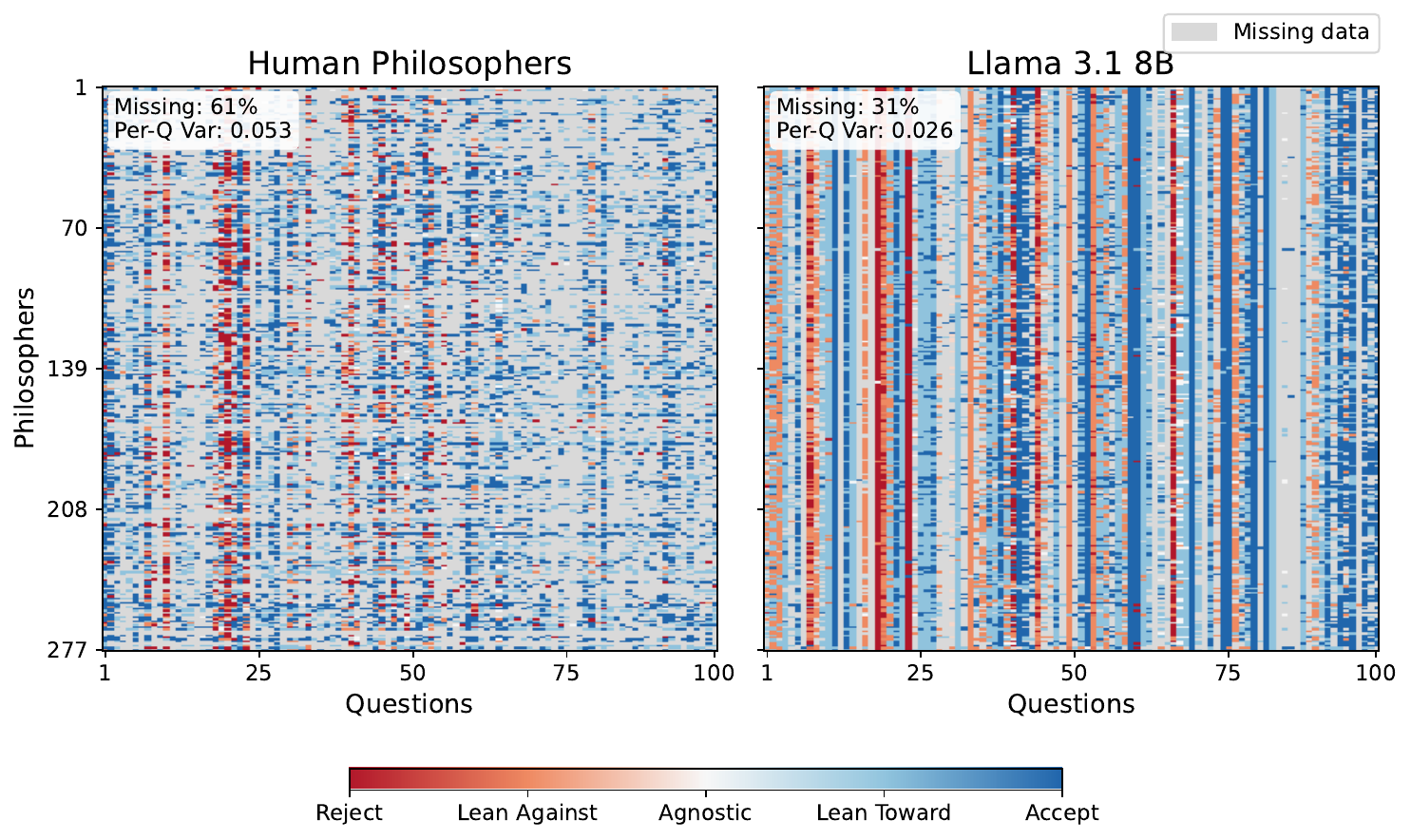}
\caption{Human vs.\ Llama-3.1-8B}
\label{fig:human_llama}
\end{subfigure}
\caption{Response matrices comparing human philosophers (left panels) and LLM simulations (right panels). Each row is a philosopher ($N=277$), each column a question ($Q = 100$). Color: red = Reject, white = Agnostic, blue = Accept; gray = missing data.}
\Description{Two vertically stacked comparisons showing response matrices. Human matrices show heterogeneous color patterns indicating diverse philosophical positions. LLM matrices show more uniform coloring, particularly visible in Llama-3.1-8B, indicating reduced within-question variance.}
\label{fig:heterogeneity}
\end{figure}

Figure~\ref{fig:heterogeneity} visualizes the core finding: heterogeneity collapse in silicon sampling. Each cell represents a philosopher's response to a question, with colors ranging from red (Reject) through white (Agnostic) to blue (Accept). Human philosophers exhibit substantial within-question variation (visible as color heterogeneity within columns), while LLM simulations produce more uniform responses. See Appendix~\ref{app:8panel} for comparisons across all seven models.

\subsection{Overall Response Matrix Similarity}
\label{sec:matrix_sim}

Before examining fine-grained structural properties, we assess how closely each LLM's full response matrix matches the human matrix. We compute KL divergence, JS divergence, and Pearson correlation on the flattened $277 \times 100$ response matrices (distinct from the correlation-distribution comparisons in Section~\ref{sec:question_corr}). Table~\ref{tab:matrix_similarity} ranks models by proximity to human responses.

\begin{table}[t]
\begin{minipage}[c]{0.52\columnwidth}
\centering
\footnotesize
\setlength{\tabcolsep}{4pt}
\begin{tabular}{@{}lccc@{}}
\toprule
Model & KL Div.\ ($\downarrow$) & JS Div.\ ($\downarrow$) & Pearson $r$ ($\uparrow$) \\
\midrule
Claude-Sonnet-4.5 & 0.052 & 0.010 & 0.495 \\
Llama-3.1-8B      & 0.055 & 0.015 & 0.268 \\
Llama-3.1-8B (FT) & 0.291 & 0.087 & 0.272 \\
Mistral-7B        & 0.653 & 0.178 & 0.264 \\
Qwen-3-4B         & 0.907 & 0.160 & 0.291 \\
GPT-5.1           & 0.974 & 0.142 & 0.422 \\
GPT-4o            & 1.012 & 0.145 & 0.416 \\
\bottomrule
\end{tabular}
\end{minipage}%
\hfill
\begin{minipage}[c]{0.44\columnwidth}
\caption{Overall response matrix similarity to human data, ranked by KL divergence. All three metrics computed on flattened response matrices using pairwise non-missing entries (both human and model response present for that philosopher--question cell). Full pairwise matrices in Appendix~\ref{app:matrix_similarity}.}
\label{tab:matrix_similarity}
\end{minipage}
\end{table}

Claude-Sonnet-4.5 is closest to human responses across all three metrics (KL$\,=\,$0.052, JS$\,=\,$0.010, $r\,=\,$0.495). GPT-5.1 and GPT-4o produce near-identical responses (JS${}<{}$0.001, $r\,=\,$0.951; \qty{95.7}{\%} of paired cells are exactly equal), possibly reflecting architectural or training overlap. DPO fine-tuning leaves Llama-3.1-8B's matrix-level correlation with human data essentially unchanged (0.268$\,\to\,$0.272) while substantially increasing JS divergence (0.015$\,\to\,$0.087): the response distribution drifts away from human even as inter-question correlation structure improves (Corr.\ $r$: 0.020$\,\to\,$0.044$^*$; see Section~\ref{sec:question_corr}). This distribution--structure split is the variance--fidelity trade-off in its clearest form.

LLMs produce correlations exceeding human magnitudes. The strongest human correlation in our sample is $|r| = 0.80$; LLMs produce perfect correlations ($|r| = 1.0$) for some specialist--question pairs due to near-uniform responses within specialist groups. LLMs also show roughly twice the rate of significant correlations (\qty{7.6}{\%}--\qty{9.0}{\%} vs.\ \qty{4.1}{\%}).

We have observed two patterns and validate each using $t$-tests across our 277-philosopher ground truth.\footnote{The matching-correlation test is a pre-registered replication of an effect reported in \citet{bourget2023philosophers}; the four amplified/spurious effects (one in prose, three in Table~\ref{tab:spurious_correlations}) have LLM effect sizes of $+43$ to $+69$ percentage points and survive Bonferroni correction for multiple testing across the demographic-correlation scan.}

\paragraph{Matching correlations.} Philosophy of Religion specialists in our ground truth endorse cosmological arguments for theism at \qty{+40.3}{\pp} (percentage points) vs.\ non-specialists ($p<0.01$), and all LLMs replicate this direction (e.g., Claude-Sonnet-4.5: \qty{+60.1}{\pp}, $p<0.001$), accurately reflecting the strong human pattern.

\paragraph{Amplified and spurious correlations.} LLMs systematically inflate weak or non-existent human correlations into strong, significant ones. Metaphysics specialists show a weak, non-significant tendency toward Humean laws of nature (\qty{-8.2}{\pp}, $n=54$), but LLMs amplify this to \qty{-19.3}{\pp} on average (four of seven models significant at $p<0.001$). In more extreme cases, correlations appear significant in LLM predictions but are entirely absent from human data. Table~\ref{tab:spurious_correlations} presents candidate spurious effects, following the two-source validation approach described in Section~\ref{sec:evaluation}.

\begin{table*}[t]
\centering
\footnotesize
\setlength{\tabcolsep}{4pt}
\begin{tabular}{@{}lcccc@{}}
\toprule
& \multicolumn{2}{c}{\emph{Ground Truth ($N=277$)}} & \multicolumn{2}{c}{\emph{LLM Predictions}} \\
\cmidrule(lr){2-3} \cmidrule(lr){4-5}
Specialist Effect & Diff & Sig & Avg Diff & Sig Models \\
\midrule
Phil.\ Biology $\rightarrow$ Personal identity: biological  & \qty{+11.4}{\pp} & n.s. & \qty{+43}{\pp} & 4/7*** \\
Phil.\ Biology $\rightarrow$ Personal identity: psychological  & \qty{+4.1}{\pp} & n.s. & \qty{-65.7}{\pp} & 3/7*** \\
Ancient Phil.\ $\rightarrow$ Practical reason: Aristotelian  & \qty{+1.9}{\pp} & n.s. & \qty{+68.9}{\pp} & 7/7*** \\
\bottomrule
\end{tabular}
\caption{Candidate spurious specialist effects. Ground truth shows no significant effect; LLMs predict highly significant associations (***$p<0.001$). Ground truth has $n<5$ specialists for these effects.}
\label{tab:spurious_correlations}
\end{table*}

The Philosophy of Biology $\rightarrow$ biological personal identity correlation illustrates this: our ground truth shows a weak, non-significant association (\qty{+11.4}{\pp}, $n=12$), the PhilPapers 2020 Survey reports no significant specialist effect for this pairing, yet four of seven LLMs predict large positive differences (average $\approx$\qty{43}{\pp}, $p<0.001$). The surface-level lexical overlap between ``biology'' and ``biological'' may explain this spurious prediction---a pattern consistent across architectures. Similarly, LLMs predict Ancient Philosophy specialists strongly endorse Aristotelian practical reason (\qty{+68.9}{\pp} average, all 7 models significant), while our ground truth shows only \qty{+1.9}{\pp} (not significant).

These examples suggest LLMs anchor on demographic labels to assign stereotypical stances, failing to recover the agnosticism and cross-domain complexity seen in human experts. The questions most predictable from demographics are listed in Appendix~\ref{app:predictable_questions}.

These findings align with \citet{santurkar2023whose}, who showed that LLM opinions systematically diverge from human populations, and \citet{durmus2023towards}, who documented biased representations of global opinions. Our results extend these concerns to expert domains.

\subsection{Question-Level Predictability Analysis}

Complementing the demographic-position analysis, we examine algorithmic fidelity from the \emph{question} perspective: which philosophical questions and domains are LLMs most and least able to predict (RQ1 continued)? This reveals where silicon sampling succeeds and where it fails, informing researchers about which philosophical topics can be reliably simulated. We compute per-question root-mean-square error (RMSE) on the normalized data (one variable per question stem, following \citeauthor{bourget2023philosophers}) against the 277-philosopher ground truth, averaged across all models.

\paragraph{Question-Level Results.} Table~\ref{tab:question_predictability} shows the most and least predictable individual questions across all 100 PhilPapers questions.

\begin{table}[t]
\begin{minipage}[c]{0.52\columnwidth}
\centering
\footnotesize
\setlength{\tabcolsep}{2pt}
\begin{tabular}{@{}lcc@{}}
\toprule
Question & RMSE & Domain \\
\midrule
\multicolumn{3}{@{}l}{\emph{Most predictable}} \\
Statue and lump: two things & 0.110 & Metaphysics \\
Other minds: adult humans & 0.114 & Phil.\ of Mind \\
Sleeping beauty: one-third & 0.132 & Decision Theory \\
Semantic content: moderate contextualism & 0.151 & Phil.\ of Language \\
Aim of philosophy: understanding & 0.164 & Phil.\ Methodology \\
\addlinespace
\multicolumn{3}{@{}l}{\emph{Least predictable}} \\
Continuum hypothesis: determinate & 0.500 & Logic \& Formal Phil. \\
Immortality: yes & 0.498 & Phil.\ of Religion \\
Extended mind: yes & 0.493 & Phil.\ of Mind \\
Arguments for theism: design & 0.489 & Phil.\ of Religion \\
Metaontology: heavyweight realism & 0.480 & Metaphysics \\
\bottomrule
\end{tabular}
\end{minipage}%
\hfill
\begin{minipage}[c]{0.44\columnwidth}
\caption{Most and least predictable individual questions (normalized following \citet{bourget2023philosophers}, one variable per question stem). RMSE ranges from 0.110 to 0.500.}
\label{tab:question_predictability}
\end{minipage}
\end{table}

LLM prediction accuracy varies by a factor of $4.6$ across questions (RMSE 0.110--0.500). However, much of this variation is explained by human ground-truth variance: per-question RMSE correlates strongly with human response variance ($r = 0.75$, $p < 0.001$). The most predictable questions have near-consensus among philosophers (e.g., ``statue and lump: two things'' and ``other minds: adult humans'' have very low human variance), while the least predictable have high variance reflecting genuine disagreement (e.g., ``immortality: yes,'' ``extended mind: yes''). This means LLMs are not selectively better at certain philosophical topics. Instead, they simply benefit from questions where there is less disagreement to capture. The least predictable questions involve contested positions where philosophers' stances are shaped by individual theoretical commitments (``metaontology: heavyweight realism,'' ``continuum hypothesis: determinate'') or personal convictions (``immortality: yes,'' ``arguments for theism: design'').

\paragraph{Domain-Level Results.} Table~\ref{tab:domain_predictability} presents philosophical domains ranked by average prediction error. We assign each of the 100 PhilPapers questions to one of 14 domains using an LLM-assisted categorization with human validation (see Appendix~\ref{app:domain_assignments} for complete assignments). Questions spanning multiple domains are assigned to their primary domain based on the question's central topic. RMSE values are computed on the normalized $[0,1]$ response scale, enabling cross-question comparison.

\begin{table}[t]
\begin{minipage}[c]{0.52\columnwidth}
\centering
\footnotesize
\setlength{\tabcolsep}{4pt}
\begin{tabular}{@{}rlccc@{}}
\toprule
Rank & Domain & RMSE & H.Var & N \\
\midrule
1 & Aesthetics & 0.216 & 0.014 & 1 \\
2 & Philosophy of Language & 0.257 & 0.036 & 7 \\
3 & History of Philosophy & 0.271 & 0.031 & 4 \\
4 & Political \& Social Phil. & 0.278 & 0.044 & 7 \\
5 & Phil.\ Methodology & 0.279 & 0.064 & 6 \\
6 & Philosophy of Mind & 0.284 & 0.044 & 14 \\
\addlinespace
7 & Metaphysics & 0.295 & 0.040 & 17 \\
8 & Decision Theory & 0.297 & 0.032 & 3 \\
9 & Epistemology & 0.310 & 0.053 & 10 \\
10 & Philosophy of Science & 0.323 & 0.032 & 4 \\
\addlinespace
11 & Logic \& Formal Phil. & 0.349 & 0.036 & 5 \\
12 & Ethics \& Moral Phil. & 0.362 & 0.082 & 10 \\
13 & Applied Ethics & 0.406 & 0.100 & 8 \\
14 & Philosophy of Religion & 0.412 & 0.097 & 4 \\
\bottomrule
\end{tabular}
\end{minipage}%
\hfill
\begin{minipage}[c]{0.44\columnwidth}
\caption{Domain predictability ranking. RMSE (lower = more predictable); H.Var = human variance; $N$ = questions per domain.}
\label{tab:domain_predictability}
\end{minipage}
\end{table}

Domain-level predictability varies by a factor of $1.9$ (RMSE 0.216--0.412). The H.Var column reveals why: normatively-loaded domains have the highest human variance (Applied Ethics 0.100, Philosophy of Religion 0.097, Ethics 0.082) and correspondingly high RMSE, while domains with lower disagreement (Aesthetics 0.014, Philosophy of Language 0.036) are most predictable. The same mechanism identified at the question level---LLMs appearing accurate where philosophers already agree. Per-model RMSE for selected domains (Appendix~\ref{app:domain_per_model}) reveals substantial model-level variation: for example, Claude-Sonnet-4.5 achieves RMSE 0.23 on Metaphysics while Mistral-7B scores 0.37. The model-level differences may reflect varying training data composition, alignment procedures, or architectural choices, though we lack direct evidence to identify specific causal factors.

\subsection{Principal Component Analysis: Structural Alignment}
\label{sec:PCA}
Following the methodology of \citet{bourget2023philosophers}, we apply PCA to characterize the dimensionality of philosophical disagreement and assess structural alignment between human and LLM responses (RQ1 continued). Table~\ref{tab:pca_comparison} summarizes the PCA results across all sources.

\begin{table}[t]
\begin{minipage}[c]{0.52\columnwidth}
\centering
\footnotesize
\setlength{\tabcolsep}{4pt}
\begin{tabular}{@{}lcccc@{}}
\toprule
Source & Var(6) & Elem $r$ & Load $r$ & Q.\ Overlap \\
\midrule
Human ($N=277$) & \qty{71.1}{\%} & --- & --- & --- \\
\midrule
Claude-Sonnet-4.5 & \qty{72.3}{\%} & 0.245** & 0.023 & 0.8/5 \\
GPT-4o & \qty{83.0}{\%} & 0.103** & 0.110 & 0.8/5 \\
GPT-5.1 & \qty{81.6}{\%} & 0.099** & $-$0.137 & 0.7/5 \\
Llama-3.1-8B (FT) & \qty{61.2}{\%} & 0.065* & $-$0.063 & 0.7/5 \\
Mistral-7B & \qty{73.8}{\%} & 0.027 & $-$0.041 & 0.7/5 \\
Qwen-3-4B & \qty{67.8}{\%} & 0.027 & 0.045 & 0.8/5 \\
Llama-3.1-8B & \qty{62.4}{\%} & 0.016 & 0.026 & 0.7/5 \\
\bottomrule
\end{tabular}
\end{minipage}%
\hfill
\begin{minipage}[c]{0.44\columnwidth}
\caption{PCA structural comparison (ranked by element-wise $r$). Var(6) = variance explained by top 6 components; Elem $r$ = element-wise correlation between question correlation matrices (higher is better); Load $r$ = correlation between flattened PCA loading matrices; Q.\ Overlap = average top-5 question overlap per component.}
\label{tab:pca_comparison}
\end{minipage}
\end{table}

\paragraph{Human PCA structure.} Human philosophers exhibit 5 principal components each explaining $\geq$\qty{2}{\%} variance, with the top 6 components explaining \qty{71.1}{\%} of total variance. PC1 (\qty{20.2}{\%}) loads most strongly on metaphilosophy: naturalism (+0.31), morality: naturalism (+0.28), and meta-ethics: moral realism ($-$0.28). PC2 (\qty{16.0}{\%}) captures extended mind (+0.41), environmental ethics ($-$0.33), and mental content: externalism (+0.31). PC3 (\qty{14.0}{\%}) loads on immortality (+0.42), law: legal positivism (+0.32), and Wittgenstein: late ($-$0.27). This multi-dimensionality indicates disagreement occurs along multiple independent axes without any single dimension dominating, which is consistent with the finding of \citet{bourget2023philosophers} that philosophical views do not organize along a single ``left-right'' spectrum. Full component interpretations are provided in Appendix~\ref{app:pca_loadings}.

\paragraph{LLM vs. Human First Principal Component (PC1) Comparison.} Human PC1 spans meta-ethics, metaphilosophy, morality, and philosophy of religion, while LLM PC1s show greater topical clustering: GPT-4o loads heavily on mind uploading and spacetime, while Llama-3.1-8B clusters external-world skepticism and theory of reference. Claude-Sonnet-4.5 shows the best PC1 alignment with human loadings ($|r|=0.63$), sharing metaphilosophy and morality: naturalism in its top loadings. Per-model PC1 loadings are reported in Appendix~\ref{app:pc1_loadings}.

\paragraph{Structural alignment metrics.} We assess structural alignment using three complementary metrics: \emph{element-wise correlation} (whether question pairs correlate similarly), \emph{loading correlation} (whether questions have similar PCA loadings), and \emph{question overlap} (whether top-loading questions match). Full metric definitions appear in Appendix~\ref{app:pca_loadings}.

Most models capture more variance in their top-6 components (GPT-4o \qty{83.0}{\%}, GPT-5.1 \qty{81.6}{\%}, Mistral-7B \qty{73.8}{\%}) than humans (\qty{71.1}{\%}), indicating LLM disagreement collapses onto fewer dimensions---consistent with heterogeneity collapse. Open-source models Llama-3.1-8B (\qty{62.4}{\%}) and Llama-3.1-8B (FT) (\qty{61.2}{\%}) capture \emph{less} variance than humans, suggesting these models distribute variance more diffusely across components. Element-wise correlations range from weak to moderate (0.016--0.245), with Claude-Sonnet-4.5 achieving the best structural alignment (0.245), followed by GPT-4o (0.103) and GPT-5.1 (0.099). Loading correlations remain weak ($-0.137$ to $0.110$). Human PC1 spans meta-ethics, metaphilosophy, and morality, while LLM PC1s cluster by surface topic (e.g., GPT-4o concentrates on mind uploading/spacetime). Per-component comparisons in Appendix~\ref{app:pca_loadings} show best individual alignment for Claude-Sonnet-4.5 PC1 ($|r| = 0.63$), with most below 0.20.

\subsection{Question Correlation}
\label{sec:question_corr}

The preceding analyses examined whether LLMs correctly predict \emph{which} positions philosophers hold and whether these predictions depend appropriately on demographic features. We now examine whether LLMs preserve the relationships \emph{between} philosophical questions (RQ2). In human data, positions on different questions are correlated: physicalists tend toward atheism ($r \approx 0.37$), and ethics questions cluster together. Table~\ref{tab:question_correlations} reports metrics assessing this structural preservation.

\begin{table}[t]
\begin{minipage}[c]{0.48\columnwidth}
\centering
\footnotesize
\setlength{\tabcolsep}{3pt}
\begin{tabular}{@{}lcccc@{}}
\toprule
Model & Elem $r$ & RV & KL & JS \\
\midrule
Sonnet-4.5 & 0.182** & 0.318 & 0.188 & 0.048 \\
GPT-5.1 & 0.089** & 0.307 & 0.468 & 0.110 \\
GPT-4o & 0.088** & 0.318 & 0.485 & 0.114 \\
Llama-3.1-8B (FT) & 0.044* & 0.224 & 0.176 & 0.046 \\
Mistral-7B & 0.043* & 0.267 & 0.387 & 0.088 \\
Qwen-3-4B & 0.024 & 0.321 & 0.670 & 0.138 \\
Llama-3.1-8B & 0.020 & 0.234 & 0.195 & 0.046 \\
\bottomrule
\end{tabular}
\end{minipage}%
\hfill
\begin{minipage}[c]{0.48\columnwidth}
\caption{Question correlation structure preservation. Elem $r$, RV, and Mantel test compare $100 \times 100$ correlation matrices; KL/JS compare correlation coefficient distributions. **$p<0.01$, *$p<0.05$.}
\label{tab:question_correlations}
\end{minipage}
\end{table}

Claude-Sonnet-4.5 leads on element-wise $r$ (0.182). The fine-tuned Llama-3.1-8B achieves the lowest KL divergence (0.176) and ties for lowest JS (0.046). GPT-5.1 and GPT-4o perform similarly on structure preservation despite different capabilities.

\subsection{The Effects of Finetuning}
Table~\ref{tab:finetuning_tradeoff} compares base and fine-tuned Llama-3.1-8B across response diversity and structural coherence metrics.

\begin{table}[t]
\begin{minipage}[c]{0.52\columnwidth}
\centering
\footnotesize
\setlength{\tabcolsep}{3pt}
\begin{tabular}{@{}lccccc@{}}
\toprule
& \multicolumn{2}{c}{\emph{Resp.\ Div.}} & \multicolumn{3}{c}{\emph{Struct.\ Coh.}} \\
\cmidrule(lr){2-3} \cmidrule(lr){4-6}
Model & Entr. & KL & Elem $r$ & C.KL & C.JS \\
\midrule
Llama Base & 0.737 & 0.07 & 0.020 & 0.195 & 0.046 \\
Llama DPO  & 0.794 & 0.40 & 0.044* & 0.176 & 0.046 \\
\bottomrule
\end{tabular}
\end{minipage}%
\hfill
\begin{minipage}[c]{0.44\columnwidth}
\caption{Fine-tuning trade-off: response diversity vs.\ structural coherence for base and DPO-tuned Llama-3.1-8B.}
\label{tab:finetuning_tradeoff}
\end{minipage}
\end{table}

The fine-tuned model better preserves inter-question correlations yet shifts the overall response distribution further from human. Corr.\ KL improves from 0.195 to 0.176 (\qty{-10}{\%}), while Corr.\ JS remains unchanged at 0.046. Element-wise correlation doubles (0.020$\to$0.044, reaching significance at $p<0.05$). Average entropy increases (0.737$\to$0.794), though the number of uniform-response questions remains similar (9 vs.\ 10). However, the overall response distribution shifts markedly: the fine-tuned model assigns \qty{34.7}{\%} of responses to ``Agnostic'' versus \qty{1.5}{\%} for the base model and \qty{1.4}{\%} for humans (Resp.\ KL increases from 0.07 to 0.40). 

Three mechanisms could explain this trade-off. First, DPO training may have overfit to the training distribution's Agnostic tendencies. Second, the limited training data (\num{3434} examples) may be insufficient to capture both structural relationships and within-question variation. Third, the contrastive DPO objective may bias toward dampened responses rather than preserving response distributions. We leave this to future work. Per-question effects are detailed in Appendix~\ref{app:finetuning_details}.

\section{Discussion}

These findings raise concerns for using silicon samples as proxies for expert judgment. LLMs organize philosophical disagreement along fundamentally different dimensions than humans, clustering questions by surface-level topic similarity rather than the cross-domain patterns observed in human data. This suggests models rely on lexical associations rather than coherent philosophical reasoning---a pattern consistent across all seven models, indicating systematic LLM properties rather than model-specific artifacts.

Amplified correlations and spurious specialist effects demonstrate that models apply stereotypes rather than learned patterns. Our findings align with the observation of \citet{wang2025llm} that LLMs flatten identity groups, here extended to professional philosophers. Domains critical for normative alignment show the most severe distortion: Applied Ethics (RMSE 0.406, H.Var 0.100) and Philosophy of Religion (RMSE 0.412, H.Var 0.097) rank 13th and 14th of 14 domains in predictability (Table~\ref{tab:domain_predictability}), with LLM variance collapsed roughly five-fold relative to human disagreement.

The stereotyping and predictability patterns reinforce each other. LLMs rely on demographic labels to generate responses (Table~\ref{tab:spurious_correlations}), but this succeeds only where human philosophers already converge. On contested questions where expert disagreement is most informative, the same mechanism produces spurious associations that do not reflect the ground truth. DPO fine-tuning partially mitigates structural mismatch (element-wise $r$ doubles) but does not resolve this dependency on consensus, and introduces its own distributional distortions (Section~\ref{sec:question_corr}).

Question-level predictability is largely explained by human consensus, not model capability: RMSE correlates strongly with ground-truth variance ($r = 0.75$). LLMs appear accurate on questions where philosophers already agree, but this masks their inability to reproduce the structured disagreement that defines the most contested philosophical positions, which are arguably the most consequential for AI alignment.

These results suggest that silicon sampling, while useful for simulating aggregate population trends \citep{argyle2023out}, may be inadequate for applications requiring faithful representation of expert disagreement. Researchers should exercise particular caution when using LLMs to simulate expert judgment in normatively-loaded domains, and explicitly measure structural alignment rather than only aggregate accuracy.

There are two areas of limitations of this paper. First, our 277-philosopher sample exhibits geographic bias, overrepresenting philosophers with North American PhDs (\qty{73.8}{\%} of the 214 respondents with known PhD country); \qty{22.7}{\%} lack PhD country information entirely. While topical diversity (AOS categories) is preserved, these geographic skews may affect generalizability, particularly for questions where regional philosophical traditions differ substantively. Other selection biases arise as philosophers maintaining active PhilPeople profiles may differ systematically (more online engagement, younger, more technologically literate). Self-reported positions may have changed since data collection, introducing temporal inconsistency. A second source arises from the high missingness of our data, and the necessity to impute values. Our principal component analysis uses iterative PCA imputation following the methodology of \citet{bourget2023philosophers} (equivalent to R's \texttt{missMDA} package). Imputation is used only for PCA; question correlation structure analysis uses pairwise-deletion correlation matrices to avoid introducing artificial structure. While imputation leverages the correlation structure to fill missing values, it necessarily introduces model-based assumptions. LLM data has lower missingness (\qty{17}{\%}--\qty{40}{\%}) than human data (\qty{61.1}{\%}), creating an asymmetry: imputation has a larger effect on human data. Our primary findings (heterogeneity collapse, failure modes) are robust to this asymmetry, as they are confirmed by multiple independent metrics (response diversity, demographic correlations, matrix similarity) that do not rely on imputation.

\section{Conclusion}

Our evaluation of seven LLMs against 277 philosophers and the PhilPapers 2020 Survey reveals three principal findings.

First, all LLMs exhibit structural mismatch in how philosophical disagreement is organized. LLMs cluster questions by surface-level topic similarity rather than the cross-domain patterns observed in human data. Proprietary models capture more variance in fewer principal components, which is consistent with heterogeneity collapse.

Second, LLMs exhibit inflated demographic-position correlations with spurious specialist effects. This ``stereotyping'' behavior---models anchoring on demographic labels to assign stereotypical stances rather than recovering expert agnosticism---explains their failure to capture the structures found in expert data.

Third, DPO fine-tuning improves correlation structure preservation but shifts the overall response distribution further from human, suggesting preference optimization faces inherent trade-offs between structural fidelity and distributional accuracy.

These results suggest that silicon sampling is inadequate for applications requiring faithful representation of expert disagreement in normatively-loaded domains. Alignment research must account for the fact that LLMs may produce artificial consensus that does not reflect the reasoned disagreement of experts.

There are several avenues for future work. First, validating against official PhilPapers individual data would strengthen claims about ground-truth patterns. Second, characterizing the temperature-fidelity trade-off would clarify whether sampling diversity mitigates heterogeneity collapse. Third, exploring alternative fine-tuning approaches (diversity-regularized objectives, varying DPO $\beta$) could determine whether the diversity-structure trade-off is inherent or can be overcome. Finally, developing metrics that explicitly quantify heterogeneity preservation alongside accuracy could improve silicon sampling evaluation.

\begin{acks}
This work is generously supported by Adobe. We thank David Bourget for agreeing to the use of PhilPapers and PhilPeople web data, Professor Sanmi Koyejo, Sang Truong, and the participants and class staff of Stanford University's CS329H \enquote{Machine Learning from Human Preferences} for helpful conversations.
\end{acks}

\bibliographystyle{ACM-Reference-Format}
\bibliography{references}

\appendix
\makeatletter
\addtocontents{toc}{\protect\etoc@depthtag{appendix}}
\makeatother

\section*{Generative AI Usage Statement}
This research leveraged Generative AI tools (Claude, ChatGPT) in several capacities. For code generation, AI assisted with implementing web scrapers (e.g., Selenium-based data collection from PhilPeople), data processing pipelines, and SVG-graphics generation; the authors carefully verified all generated code, conducted debugging, and took full responsibility for its correctness. For writing, AI helped identify grammatical errors, improve sentence structure, and polish academic tone. For methodology, AI suggested relevant statistical metrics (e.g., the Mantel test, RV coefficient) that were independently verified against statistical literature before adoption. The authors maintain full ownership of research design, hypothesis formulation, experimental execution, result interpretation, and all conclusions drawn.

\section*{Appendices}

\begingroup
\etocsettagdepth{appendix}{subsection}
\etocsettagdepth{mainpaper}{none}
\etocsetnexttocdepth{subsection}
\etocsettocstyle{\section*{}}{}
\tableofcontents
\endgroup

\section{Data}

\subsection{Demographic Feature Categories}
\label{app:demographic_features}

Demographic features used in correlation analyses include four categories:

\paragraph{Areas of Specialization (AOS) and Areas of Interest (AOI)} Ancient Greek and Roman Philosophy, 17th/18th Century Philosophy, 19th Century Philosophy, 20th Century Philosophy, Applied Ethics, Epistemology, General Philosophy of Science, Logic and Philosophy of Logic, Meta-Ethics, Metaphysics, Normative Ethics, Philosophy of Action, Philosophy of Biology, Philosophy of Cognitive Science, Philosophy of Language, Philosophy of Mind, Philosophy of Physical Science, Philosophy of Religion, and Social and Political Philosophy

\paragraph{PhD Country.} USA, United Kingdom, Canada, Germany, Australia, and others (including ``Unknown'' for missing data).

\paragraph{PhD Year.} Binned by 5-year intervals (e.g., 1990--1994, 1995--1999, 2000--2004).

Demographic features (AOS, AOI, PhD country, PhD-year bins) are tested at the (feature, question) pair level: each pair contributes a test when the question has at least six valid responses and the feature varies among respondents (i.e., at least one philosopher has the feature and at least one does not). This yields approximately 400 effective features per model that contribute at least one test.

\subsection{Question-Domain Assignments}
\label{app:domain_assignments}

We assign all 100 PhilPapers questions to one of 14 philosophical domains using LLM-assisted categorization (Claude-Opus-4.5) with human validation. Domains are ordered by predictability (RMSE), from most predictable (Aesthetics) to least predictable (Philosophy of Religion).

\begin{description}[style=nextline, font=\normalfont\itshape, leftmargin=1em]
\item[Aesthetics (1)] aesthetic value
\item[Philosophy of Language (7)] analytic-synthetic distinction, proper names, propositions, semantic content, theory of reference, truth, vagueness
\item[History of Philosophy (4)] hume, kant, plato, wittgenstein
\item[Political \& Social Phil.\ (7)] gender, gender categories, law, political philosophy, politics, race, race categories
\item[Phil.\ Methodology (6)] aim of philosophy, metaphilosophy, method in history of philosophy, method in political philosophy, philosophical methods, philosophical progress
\item[Philosophy of Mind (14)] aesthetic experience, chinese room, concepts, consciousness, extended mind, grounds of intentionality, hard problem of consciousness, mental content, mind, mind uploading, other minds, perceptual experience, propositional attitudes, zombies
\item[Metaphysics (17)] abstract objects, causation, external world, free will, interlevel metaphysics, laws of nature, material composition, metaontology, personal identity, possible worlds, properties, spacetime, statue and lump, teletransporter, temporal ontology, time, time travel
\item[Decision Theory (3)] newcomb's problem, practical reason, sleeping beauty
\item[Epistemology (10)] a priori knowledge, analysis of knowledge, belief or credence, epistemic justification, justification, knowledge, knowledge claims, philosophical knowledge, rational disagreement, response to external-world skepticism
\item[Philosophy of Science (4)] quantum mechanics, science, units of selection, values in science
\item[Logic \& Formal Phil.\ (5)] continuum hypothesis, foundations of mathematics, logic, principle of sufficient reason, true contradictions
\item[Ethics \& Moral Phil.\ (10)] meaning of life, meta-ethics, moral judgment, moral motivation, moral principles, morality, normative concepts, normative ethics, ought implies can, well-being
\item[Applied Ethics (8)] abortion, capital punishment, eating animals and animal products, environmental ethics, experience machine, footbridge, human genetic engineering, trolley problem
\item[Philosophy of Religion (4)] arguments for theism, cosmological fine-tuning, god, immortality
\end{description}

\section{Language Models}

\subsection{Prompt Templates}
\label{app:prompts}

\paragraph{Persona Prompt.} The persona prompt encodes the philosopher's professional identity:
\begin{verbatim}
You are a professional philosopher at {institution} ({country}). Your Educational Background:
- PhD from {phd_institution} ({phd_country}) in {phd_year}
Your Areas of Specialization:
- {specialization_1} ... [up to 5]
Your Areas of Interest:
- {interest_1} ... [up to 8]
\end{verbatim}

\paragraph{Question Prompt.}

\begin{verbatim}
You are answering a survey by the reputable website
PhilPapers, which collects responses across various
philosophical domains based on your background. You may
select one or multiple options if you deem it appropriate
to accurately represent your position.

Please respond with your chosen option(s) as a JSON list.
If selecting multiple options, ensure they are logically
consistent.

Examples of valid responses:
- ["Accept: physicalism"]
- ["Accept: physicalism", "Reject: non-physicalism"]
- ["Lean towards: physicalism",
   "Lean against: non-physicalism"]

Given your philosophical profile above, please express
your stance on the following question.

Question: {question_key}

Available response options:
- {option_1}
- {option_2}
...

Respond ONLY with the JSON list, no other text or
explanation.
\end{verbatim}

\subsection{Response Parsing}
\label{app:parsing}

Table~\ref{tab:success_rates} reports parsing success rates.

\begin{table}[t]
\begin{minipage}[c]{0.52\columnwidth}
\centering
\footnotesize
\setlength{\tabcolsep}{4pt}
\begin{tabular}{@{}lcc@{}}
\toprule
Model & Success Rate & Primary Failure Mode \\
\midrule
GPT-5.1 & \qty{100.0}{\%} & --- \\
GPT-4o & \qty{100.0}{\%} & --- \\
Claude-Sonnet-4.5 & \qty{100.0}{\%} & --- \\
\midrule
Qwen-3-4B & \qty{99.9}{\%} & Invalid option format \\
Mistral-7B & \qty{98.1}{\%} & Invalid option format \\
Llama-3.1-8B & \qty{93.9}{\%} & ``Combination of views'' \\
Llama-3.1-8B (FT) & \qty{89.4}{\%} & ``Combination of views'' \\
\bottomrule
\end{tabular}
\end{minipage}%
\hfill
\begin{minipage}[c]{0.44\columnwidth}
\caption{Response parsing success rates.}
\label{tab:success_rates}
\end{minipage}
\end{table}

Analysis of Llama-3.1-8B failures reveals that the majority (\qty{66.3}{\%}) involve generating hedging responses not in the valid option set. DPO fine-tuning increased failures from \qty{6.1}{\%} to \qty{10.6}{\%}, with ``combination of views'' failures rising from \num{686} to \num{2312}.

\subsection{Prompt Sensitivity}
\label{app:prompt_sensitivity}

To assess whether our findings depend on prompt wording, we conduct a prompt sensitivity study. We select one proprietary model (Claude-Sonnet-4.5) and one open-source model (Llama-3.1-8B) as representatives, and re-query each on a random \qty{10}{\%} subsample of philosophers ($n=27$, seed=42) across all 100 questions using an alternative framing that removes the PhilPapers institutional context. The baseline prompt begins ``You are answering a survey by the reputable website PhilPapers\ldots''; the variant uses ``Based on your philosophical expertise and training, what is your considered position on the following question?'' The persona prompt (demographic conditioning) is held constant. Llama-3.1-8B was queried via the Fireworks AI batch inference API \citep{fireworks2025}.

We compare variant responses against baseline on matched (philosopher, question) pairs, computing: (1) paired Pearson $r$ (higher = more stable), (2) mean absolute difference (MAD), (3) root-mean-square error (RMSE), and (4) response flip rate (proportion shifting by $>$0.25 on $[0,1]$). Table~\ref{tab:prompt_sensitivity} reports results.

\begin{table}[t]
\begin{minipage}[c]{0.52\columnwidth}
\centering
\footnotesize
\setlength{\tabcolsep}{3pt}
\begin{tabular}{@{}lccccc@{}}
\toprule
Model & $r$ & MAD & RMSE & Flip\% & Shift \\
\midrule
Sonnet-4.5 & 0.483 & 0.087 & 0.188 & \qty{5.3}{\%} & $-$0.041 \\
Llama-3.1-8B   & 0.465 & 0.143 & 0.235 & \qty{12.2}{\%} & $+$0.039 \\
\bottomrule
\end{tabular}
\end{minipage}%
\hfill
\begin{minipage}[c]{0.44\columnwidth}
\caption{Prompt sensitivity: baseline (PhilPapers framing) vs.\ variant (direct question) on 27 philosophers $\times$ 100 questions.}
\label{tab:prompt_sensitivity}
\end{minipage}
\end{table}

Both models show moderate robustness to prompt framing (paired $r \geq 0.46$), with small mean shifts ($|{\Delta}| \leq 0.041$) that do not systematically favor one direction. Claude-Sonnet-4.5 is more stable (\qty{5.3}{\%} flip rate vs.\ \qty{12.2}{\%} for Llama-3.1-8B), consistent with the general pattern that larger proprietary models are less prompt-sensitive. The most sensitive questions differ between models: Llama-3.1-8B on \emph{Human genetic engineering} (24 flips) and \emph{Plato} (20 flips); Claude-Sonnet-4.5 on \emph{Capital punishment} (13 flips) and \emph{True contradictions} (13 flips). This question-specificity suggests prompt sensitivity reflects model-specific uncertainty rather than systematic prompt dependence. The key structural findings of the paper---heterogeneity collapse, inflated demographic correlations, and structural mismatch---are consistent across both models and both prompt framings, and are therefore unlikely to be prompt artifacts.

\section{Additional Analysis}

\subsection{Question Predictability}
\label{app:predictable_questions}

Table~\ref{tab:predictable_questions} presents the top-6 most predictable questions from demographic features for each model. Across all seven models, top-predictable pairs concentrate on questions whose topic overlaps lexically or semantically with the predicting demographic label (e.g., Phil.\ of Religion $\rightarrow$ God: atheism / arguments for theism; Aesthetics $\rightarrow$ aesthetic experience; Phil.\ of Gender $\rightarrow$ gender: social), reinforcing the stereotyping pattern identified in Table~\ref{tab:spurious_correlations}.

\begin{table}[t]
\begin{minipage}[c]{0.52\columnwidth}
\centering
\footnotesize
\setlength{\tabcolsep}{3pt}
\begin{tabular}{@{}llc@{}}
\toprule
Question & Predictor & $|r|$ \\
\midrule
\multicolumn{3}{@{}l}{{GPT-5.1}} \\
God: atheism & AOS: Phil. of Religion & 0.880 \\
Method in history & AOS: Phil. of Language & 0.799 \\
God: atheism & AOI: Phil. of Religion & 0.677 \\
God: atheism & AOS: 17th/18th C. Phil. & 0.677 \\
Foundations of math & AOS: General Phil. of Science & 0.586 \\
Analysis of knowledge & AOS: Ancient Phil. & 0.576 \\
\midrule
\multicolumn{3}{@{}l}{{GPT-4o}} \\
Analysis of knowledge & AOS: Ancient Phil. & 0.573 \\
Teletransporter & AOS: Medieval Phil. & 0.565 \\
Args for theism & AOI: Phil. of Religion & 0.544 \\
Mind: physicalism & AOS: Phil. of Religion & 0.527 \\
Normative ethics & AOS: Ancient Phil. & 0.526 \\
Propositional attitudes & Year: 2015--2019 & 0.507 \\
\midrule
\multicolumn{3}{@{}l}{{Claude-Sonnet-4.5}} \\
Well-being & AOS: Phil. of Mind & 0.651 \\
Concepts: empiricism & PhD: USA & 0.606 \\
Args for theism & AOS: Phil. of Religion & 0.603 \\
Spacetime & AOI: History of Western Phil. & 0.549 \\
Well-being & AOS: Phil. of Cog. Science & 0.533 \\
Material composition & Year: 2015--2019 & 0.515 \\
\midrule
\multicolumn{3}{@{}l}{{Llama-3.1-8B}} \\
Args for theism & AOS: Phil. of Religion & 0.788 \\
Aesthetic experience & AOS: Aesthetics & 0.685 \\
Time travel & AOI: Metaphysics & 0.564 \\
Plato & AOI: History of Western Phil. & 0.561 \\
Args for theism & AOI: Phil. of Religion & 0.554 \\
Plato & AOS: Ancient Phil. & 0.544 \\
\midrule
\multicolumn{3}{@{}l}{{Mistral-7B}} \\
Gender: social & AOI: Phil. of Gender & 0.781 \\
Gender: social & AOS: Phil. of Gender & 0.721 \\
Foundations of math & AOS: Phil. of Mathematics & 0.670 \\
Aesthetic value & AOS: Aesthetics & 0.665 \\
Political philosophy & AOS: Social \& Political Phil. & 0.663 \\
Race categories & AOI: Phil. of Gender & 0.634 \\
\midrule
\multicolumn{3}{@{}l}{{Qwen-3-4B}} \\
God: atheism & AOS: Medieval Phil. & 0.677 \\
Args for theism & AOS: Medieval Phil. & 0.665 \\
Args for theism & AOI: Medieval Phil. & 0.606 \\
God: atheism & AOI: Medieval Phil. & 0.524 \\
Politics: socialism & AOS: Social \& Political Phil. & 0.504 \\
Politics: socialism & AOI: Social \& Political Phil. & 0.499 \\
\bottomrule
\end{tabular}
\end{minipage}%
\hfill
\begin{minipage}[c]{0.44\columnwidth}
\caption{Most predictable questions from demographics (top 6 per model).}
\label{tab:predictable_questions}
\end{minipage}
\end{table}

\subsection{Domain Predictability}
\label{app:domain_per_model}

Table~\ref{tab:domain_per_model} reports per-model RMSE for selected philosophical domains, revealing how prediction difficulty varies across models.

\begin{table}[t]
\centering
\footnotesize
\setlength{\tabcolsep}{3pt}
\begin{tabular}{@{}lccccccc@{}}
\toprule
Domain & GPT-5.1 & GPT-4o & Sonnet-4.5 & Llama-3.1-8B & Llama-3.1-8B-FT & Mistral-7B & Qwen-3-4B \\
\midrule
Phil.\ of Science & 0.34 & 0.34 & 0.32 & 0.36 & 0.26 & 0.30 & 0.39 \\
Phil.\ of Mind & 0.25 & 0.25 & 0.26 & 0.27 & 0.28 & 0.37 & 0.32 \\
Metaphysics & 0.29 & 0.29 & 0.23 & 0.27 & 0.29 & 0.37 & 0.32 \\
Applied Ethics & 0.38 & 0.38 & 0.35 & 0.48 & 0.49 & 0.42 & 0.34 \\
Logic \& Formal & 0.34 & 0.29 & 0.25 & 0.31 & 0.32 & 0.36 & 0.36 \\
Phil.\ of Religion & 0.38 & 0.38 & 0.39 & 0.54 & 0.42 & 0.45 & 0.36 \\
\bottomrule
\end{tabular}
\caption{Per-model RMSE by domain (selected domains).}
\label{tab:domain_per_model}
\end{table}




\subsection{Principal Components}
\label{app:pca_loadings}

This appendix presents detailed PCA results following the methodology of \citet{bourget2023philosophers}: iterative PCA imputation ($\mathrm{ncp}=5$), eigendecomposition, and \qty{2}{\%} variance threshold for significant components. The human sample ($N=277$) has \qty{71.1}{\%} variance explained by the top 6 components (5 components $\geq$\qty{2}{\%}).

\subsubsection{Human Principal Components}
Table~\ref{tab:pc_loadings} presents the top loadings for each component.

PC1 (\qty{20.2}{\%}) captures the naturalism--moral realism axis; PC2 (\qty{16.0}{\%}) externalism and applied ethics; PC3 (\qty{14.0}{\%}) religion and normativity; PC4 (\qty{10.7}{\%}) moral motivation; PC5 (\qty{8.3}{\%}) the a priori--analytic cluster. Full loadings appear in Table~\ref{tab:pc_loadings}.

\subsubsection{Per-Model PC1 Loadings}
\label{app:pc1_loadings}

Table~\ref{tab:pc1_all} lists the top-5 PC1 loadings for human data and four representative models, chosen to span the observed range from closest structural alignment (Claude-Sonnet-4.5) to most extreme compression onto a single question (Qwen-3-4B). PC1 loadings for the remaining three models (GPT-5.1, Llama-3.1-8B FT, Mistral-7B) appear in the replication output of \texttt{recompute\_all\_tables.py}.

\begin{table}[t]
\begin{minipage}[c]{0.52\columnwidth}
\centering
\footnotesize
\setlength{\tabcolsep}{3pt}
\begin{tabular}{@{}lc@{}}
\toprule
{Human (\qty{20.2}{\%})} & \\
\midrule
Metaphilosophy & +0.31 \\
Args for theism & $-$0.29 \\
Morality & +0.27 \\
Meta-ethics & $-$0.27 \\
Hard problem & $-$0.23 \\
\midrule
{Claude-Sonnet-4.5 (\qty{31.5}{\%})} & \\
\midrule
Practical reason & +0.27 \\
Metaphilosophy & +0.25 \\
Knowledge & +0.25 \\
Epist.\ justification & +0.24 \\
Morality & +0.23 \\
\midrule
{GPT-4o (\qty{26.6}{\%})} & \\
\midrule
Mind uploading & +0.56 \\
Teletransporter & +0.32 \\
Spacetime & +0.30 \\
Mind: physicalism & $-$0.27 \\
Normative ethics & $-$0.25 \\
\midrule
{Llama-3.1-8B (\qty{18.1}{\%})} & \\
\midrule
Ext.-world skepticism & +0.41 \\
Theory of reference & +0.35 \\
God: atheism & +0.30 \\
Law & $-$0.28 \\
Meta-ethics & $-$0.25 \\
\midrule
{Qwen-3-4B (\qty{27.2}{\%})} & \\
\midrule
Analysis of knowledge & +0.88 \\
Mental content & $-$0.22 \\
Extended mind & $-$0.16 \\
Immortality & +0.16 \\
Laws of nature & +0.16 \\
\bottomrule
\end{tabular}
\end{minipage}%
\hfill
\begin{minipage}[c]{0.44\columnwidth}
\caption{PC1 top-5 loadings for human data and four representative models, each block labelled with the share of variance that PC1 explains. PCA sign is arbitrary; absolute magnitudes indicate loading strength. Loadings for the remaining three models (GPT-5.1, Llama-3.1-8B FT, Mistral-7B) are in the replication output of \texttt{recompute\_all\_tables.py}.}
\label{tab:pc1_all}
\end{minipage}
\end{table}

\subsubsection{Per-Model Component Comparison}
Table~\ref{tab:model_pc_comparison} shows per-component loading correlations between each model and human data. Most correlations are weak ($|r| < 0.2$), with the best alignment for Claude-Sonnet-4.5 PC1 (0.63).

Human components span multiple philosophical domains, while LLM components tend to cluster by surface topic similarity rather than cross-domain patterns.

\begin{table}[t]
\begin{minipage}[c]{0.52\columnwidth}
\centering
\footnotesize
\setlength{\tabcolsep}{2pt}
\begin{tabular}{@{}clc@{}}
\toprule
PC & Top Questions (by absolute loading) & Loading \\
\midrule
\multirow{5}{*}{PC1} & Metaphilosophy: naturalism & +0.31 \\
& Morality: naturalism & +0.28 \\
& Meta-ethics: moral realism & $-$0.28 \\
& Arguments for theism: design & $-$0.27 \\
& Laws of nature: Humean & +0.23 \\
\midrule
\multirow{5}{*}{PC2} & Extended mind: yes & +0.41 \\
& Environmental ethics: anthropocentric & $-$0.33 \\
& Mental content: externalism & +0.31 \\
& Epistemic justification: externalism & +0.31 \\
& Footbridge: push & $-$0.24 \\
\midrule
\multirow{5}{*}{PC3} & Immortality: yes & +0.42 \\
& Law: legal positivism & +0.32 \\
& Wittgenstein: late & $-$0.27 \\
& Logic: classical & +0.25 \\
& Normative ethics: Consequentialism & +0.23 \\
\midrule
\multirow{5}{*}{PC4} & Moral motivation: externalism & +0.37 \\
& Footbridge: push & +0.31 \\
& Rational disagreement: permissivism & $-$0.29 \\
& Laws of nature: Humean & $-$0.24 \\
& Environmental ethics: anthropocentric & $-$0.24 \\
\midrule
\multirow{5}{*}{PC5} & Analytic-synthetic distinction: yes & +0.41 \\
& A priori knowledge: yes & +0.33 \\
& Immortality: yes & $-$0.30 \\
& Meta-ethics: moral realism & $-$0.25 \\
& Properties: tropes & $-$0.25 \\
\bottomrule
\end{tabular}
\end{minipage}%
\hfill
\begin{minipage}[c]{0.44\columnwidth}
\caption{Top loadings for the five human principal components exceeding the \qty{2}{\%} variance threshold ($N=277$).}
\label{tab:pc_loadings}
\end{minipage}
\end{table}

\begin{table}[t]
\begin{minipage}[c]{0.52\columnwidth}
\centering
\footnotesize
\setlength{\tabcolsep}{3pt}
\begin{tabular}{@{}lcccccc@{}}
\toprule
Model & PC1 & PC2 & PC3 & PC4 & PC5 & PC6 \\
\midrule
GPT-5.1 & 0.39 & 0.02 & 0.10 & 0.06 & 0.04 & 0.20 \\
GPT-4o & 0.14 & 0.18 & 0.13 & 0.08 & 0.17 & 0.23 \\
Claude-Sonnet-4.5 & 0.63 & 0.20 & 0.02 & 0.15 & 0.07 & 0.05 \\
Llama-3.1-8B & 0.02 & 0.02 & 0.07 & 0.01 & 0.12 & 0.04 \\
Llama-3.1-8B (FT) & 0.19 & 0.02 & 0.10 & 0.11 & 0.17 & 0.18 \\
Mistral-7B & 0.06 & 0.23 & 0.03 & 0.10 & 0.22 & 0.14 \\
Qwen-3-4B & 0.09 & 0.12 & 0.00 & 0.13 & 0.06 & 0.30 \\
\bottomrule
\end{tabular}
\end{minipage}%
\hfill
\begin{minipage}[c]{0.44\columnwidth}
\caption{Per-component loading correlations ($|r|$) between human and each model.}
\label{tab:model_pc_comparison}
\end{minipage}
\end{table}

These results reveal that philosophical disagreement is multi-dimensional, with no single dimension dominating (PC1 explains \qty{20.2}{\%}). The top 6 components together explain \qty{71.1}{\%} of variance. LLM responses show variable compressibility (\qty{61}{\%}--\qty{83}{\%} variance in top-6), with proprietary models (GPT-5.1 \qty{81.6}{\%}, GPT-4o \qty{83.0}{\%}) compressing more than humans, consistent with heterogeneity collapse.

\subsection{Response Matrices}
\label{app:8panel}

\begin{figure*}[t]
\centering
\includegraphics[width=\textwidth]{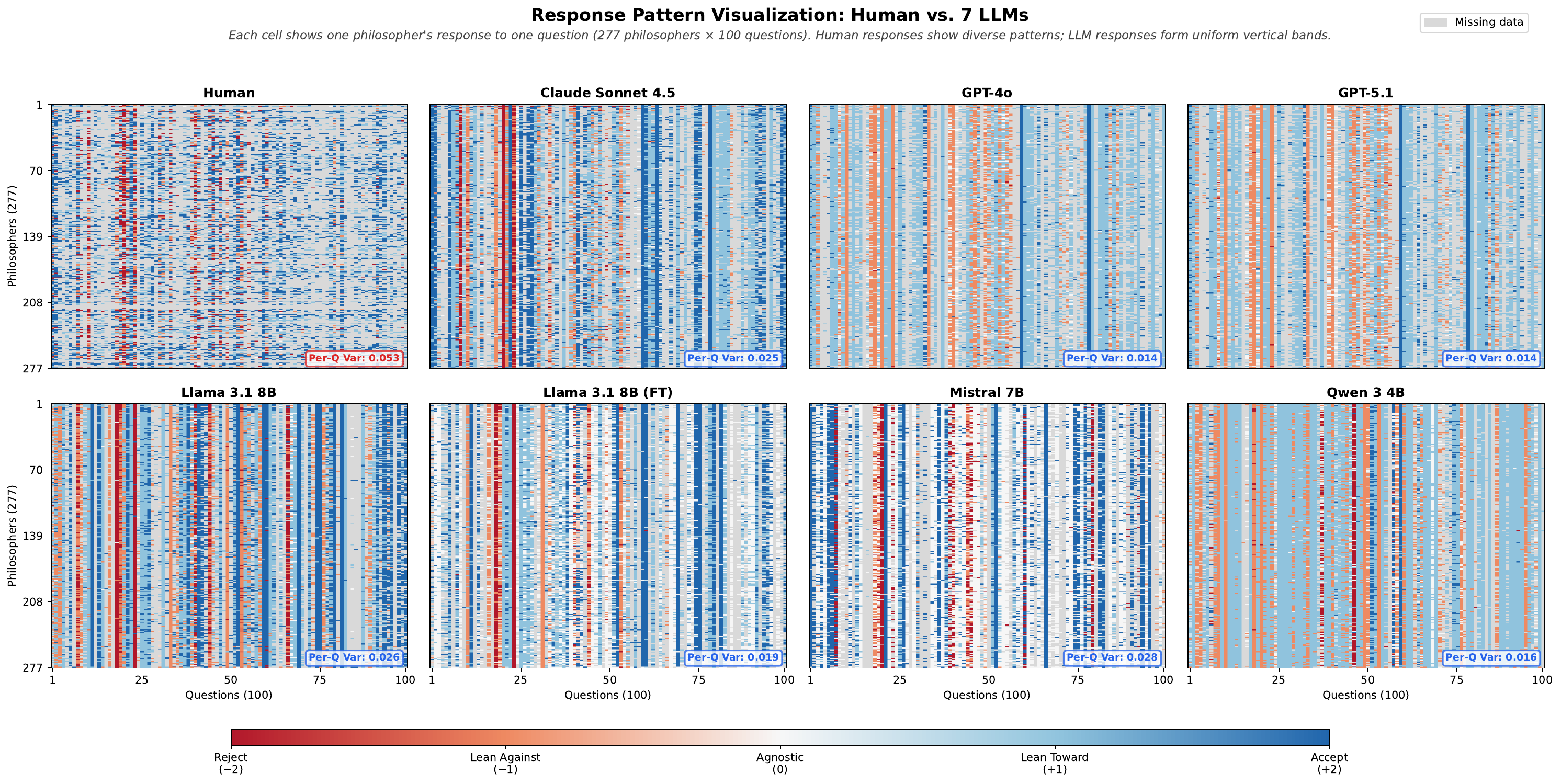}
\caption{Response matrices for human philosophers and all seven LLM simulations. Each panel: philosophers (rows) and 100 questions (columns); GPT-5.1 has 276 rows. Color scale as in Figure~\ref{fig:heterogeneity}.}
\Description{An 8-panel figure showing response matrices for human philosophers and seven LLMs. Human matrix shows diverse coloring. LLM matrices show progressively more uniform responses, with variance statistics annotated on each panel.}
\label{fig:8panel}
\end{figure*}

Figure~\ref{fig:8panel} presents response matrices for all eight data sources (human + 7 LLMs). Each panel shows 277 philosophers and 100 questions (276 for GPT-5.1), with color indicating response (red = Reject, white = Agnostic, blue = Accept) and gray indicating missing data.

\subsection{Response Matrix Similarity}
\label{app:matrix_similarity}

Tables~\ref{tab:kl_full},~\ref{tab:js_full}, and~\ref{tab:corr_full} report pairwise KL divergence, JS divergence, and Pearson correlation computed on flattened response matrices (non-missing entries only). These element-wise comparisons complement the correlation-distribution metrics in Table~\ref{tab:question_correlations} by measuring direct response overlap rather than structural similarity.

\begin{table}[t]
\begin{subtable}{\textwidth}
\centering
\footnotesize
\setlength{\tabcolsep}{3pt}
\begin{tabular}{@{}lcccccccc@{}}
\toprule
$D_{\text{KL}}(\text{row} \| \text{col})$ & Human & GPT-5.1 & GPT-4o & Sonnet-4.5 & Llama-3.1-8B & Llama-3.1-8B-FT & Mistral-7B & Qwen-3-4B \\
\midrule
Human              & ---   & 0.895 & 0.928 & 0.065 & 0.070 & 0.397 & 0.766 & 0.920 \\
\addlinespace
GPT-5.1            & 0.513 & ---   & 0.000 & 0.277 & 0.331 & 0.562 & 1.506 & 0.048 \\
GPT-4o             & 0.521 & 0.000 & ---   & 0.292 & 0.335 & 0.567 & 1.516 & 0.048 \\
Sonnet-4.5         & 0.060 & 0.438 & 0.458 & ---   & 0.046 & 0.324 & 0.893 & 0.508 \\
\addlinespace
Llama-3.1-8B       & 0.125 & 0.585 & 0.603 & 0.115 & ---   & 0.409 & 0.915 & 0.831 \\
Llama-3.1-8B-FT    & 0.874 & 0.807 & 0.820 & 0.943 & 0.876 & ---   & 0.239 & 0.915 \\
\addlinespace
Mistral-7B         & 1.596 & 1.746 & 1.773 & 1.496 & 1.733 & 0.245 & ---   & 1.875 \\
Qwen-3-4B          & 0.577 & 0.121 & 0.124 & 0.335 & 0.380 & 0.619 & 1.610 & ---   \\
\bottomrule
\end{tabular}
\caption{KL divergence (asymmetric).}
\label{tab:kl_full}
\end{subtable}

\vspace{1em}

\begin{subtable}{\textwidth}
\centering
\footnotesize
\setlength{\tabcolsep}{3pt}
\begin{tabular}{@{}lcccccccc@{}}
\toprule
JS & Human & GPT-5.1 & GPT-4o & Sonnet-4.5 & Llama-3.1-8B & Llama-3.1-8B-FT & Mistral-7B & Qwen-3-4B \\
\midrule
Human              & ---   & 0.142 & 0.144 & 0.014 & 0.019 & 0.119 & 0.214 & 0.158 \\
\addlinespace
GPT-5.1            & 0.142 & ---   & 0.000 & 0.073 & 0.094 & 0.150 & 0.332 & 0.013 \\
GPT-4o             & 0.144 & 0.000 & ---   & 0.075 & 0.095 & 0.151 & 0.334 & 0.013 \\
Sonnet-4.5         & 0.014 & 0.073 & 0.075 & ---   & 0.012 & 0.097 & 0.223 & 0.088 \\
\addlinespace
Llama-3.1-8B       & 0.019 & 0.094 & 0.095 & 0.012 & ---   & 0.117 & 0.237 & 0.107 \\
Llama-3.1-8B-FT    & 0.119 & 0.150 & 0.151 & 0.097 & 0.117 & ---   & 0.052 & 0.168 \\
\addlinespace
Mistral-7B         & 0.214 & 0.332 & 0.334 & 0.223 & 0.237 & 0.052 & ---   & 0.357 \\
Qwen-3-4B          & 0.158 & 0.013 & 0.013 & 0.088 & 0.107 & 0.168 & 0.357 & ---   \\
\bottomrule
\end{tabular}
\caption{Jensen-Shannon divergence (symmetric).}
\label{tab:js_full}
\end{subtable}

\vspace{1em}

\begin{subtable}{\textwidth}
\centering
\footnotesize
\setlength{\tabcolsep}{3pt}
\begin{tabular}{@{}lcccccccc@{}}
\toprule
Pearson $r$ & Human & GPT-5.1 & GPT-4o & Sonnet-4.5 & Llama-3.1-8B & Llama-3.1-8B-FT & Mistral-7B & Qwen-3-4B \\
\midrule
Human              & ---   & 0.397 & 0.410 & 0.445 & 0.277 & 0.257 & 0.235 & 0.274 \\
\addlinespace
GPT-5.1            & 0.397 & ---   & 0.951 & 0.691 & 0.459 & 0.495 & 0.377 & 0.505 \\
GPT-4o             & 0.410 & 0.951 & ---   & 0.731 & 0.473 & 0.512 & 0.384 & 0.518 \\
Sonnet-4.5         & 0.445 & 0.691 & 0.731 & ---   & 0.495 & 0.485 & 0.329 & 0.499 \\
\addlinespace
Llama-3.1-8B       & 0.277 & 0.459 & 0.473 & 0.495 & ---   & 0.700 & 0.314 & 0.399 \\
Llama-3.1-8B-FT    & 0.257 & 0.495 & 0.512 & 0.485 & 0.700 & ---   & 0.348 & 0.331 \\
\addlinespace
Mistral-7B         & 0.235 & 0.377 & 0.384 & 0.329 & 0.314 & 0.348 & ---   & 0.247 \\
Qwen-3-4B          & 0.274 & 0.505 & 0.518 & 0.499 & 0.399 & 0.331 & 0.247 & ---   \\
\bottomrule
\end{tabular}
\caption{Pearson correlation.}
\label{tab:corr_full}
\end{subtable}
\caption{Pairwise similarity on flattened response matrices.}
\label{tab:pairwise_full}
\end{table}

\subsection{Finetuning}

\subsubsection{Question-Level Effects}
\label{app:finetuning_details}

Table~\ref{tab:finetuning_effects} presents the questions most affected by DPO fine-tuning on Llama-3.1-8B, measured by change in Shannon entropy. Fine-tuning produces heterogeneous effects: some questions show increased diversity while others collapse to near-uniform responses.

\begin{table}[t]
\begin{minipage}[c]{0.52\columnwidth}
\centering
\footnotesize
\setlength{\tabcolsep}{3pt}
\begin{tabular}{@{}lcccc@{}}
\toprule
Question & Base & FT & $\Delta$H & $\Delta$\% \\
\midrule
\multicolumn{5}{@{}l}{\emph{Increased diversity}} \\
Immortality & 0.03 & 1.48 & +1.45 & +\num{4155} \\
Statue and lump & 0.00 & 1.00 & +1.00 & --- \\
Belief or credence & 0.26 & 1.20 & +0.94 & +357 \\
Semantic content & 0.06 & 0.93 & +0.86 & +\num{1400} \\
Perceptual exp. & 0.12 & 0.96 & +0.83 & +671 \\
\midrule
\multicolumn{5}{@{}l}{\emph{Decreased diversity}} \\
Found.\ of math & 1.56 & 0.00 & $-$1.56 & $-$100 \\
Spacetime & 1.33 & 0.00 & $-$1.33 & $-$100 \\
Law & 1.62 & 0.34 & $-$1.27 & $-$79 \\
True contradictions & 1.24 & 0.00 & $-$1.24 & $-$100 \\
Political phil. & 1.37 & 0.37 & $-$1.00 & $-$73 \\
\bottomrule
\end{tabular}
\end{minipage}%
\hfill
\begin{minipage}[c]{0.44\columnwidth}
\caption{Questions most affected by DPO fine-tuning (top 5 by absolute $\Delta$H). H = Shannon entropy. Three decreased-diversity questions collapsed to H = 0.}
\label{tab:finetuning_effects}
\end{minipage}
\end{table}

\subsubsection{SFT Finetuning}
\label{app:sft_vs_dpo}

To validate our choice of DPO for fine-tuning, we compare SFT against DPO on the same Llama-3.1-8B base model. Both methods use the PhilPapers 2009 Survey \citep{bourget2014philosophers} as their training source. SFT trains on the chosen response as a standard language modeling target; DPO uses the contrastive preference objective over chosen/rejected pairs. SFT uses LoRA ($r=16$, $\alpha=32$) with 3 epochs and learning rate \num{2e-5}; DPO uses the configuration in Section~\ref{sec:finetuning}. To facilitate execution, SFT fine-tuning was performed via the Fireworks AI platform \citep{fireworks2025}.

Table~\ref{tab:sft_vs_dpo} compares the two approaches against the base model across response diversity and structural coherence metrics.

\begin{table}[t]
\begin{minipage}[c]{0.52\columnwidth}
\centering
\footnotesize
\setlength{\tabcolsep}{3pt}
\begin{tabular}{@{}lcccccc@{}}
\toprule
& \multicolumn{3}{c}{\emph{Response Diversity}} & \multicolumn{3}{c}{\emph{Structural Coherence}} \\
\cmidrule(lr){2-4} \cmidrule(lr){5-7}
Model & Entr.$\uparrow$ & Var$\uparrow$ & Eff.Cat$\uparrow$ & Mantel $r$$\uparrow$ & Elem $r$$\uparrow$ & RV$\uparrow$ \\
\midrule
Human       & 1.293 & 0.053 & 2.56 & --- & --- & --- \\
Llama Base  & 0.770 & 0.026 & 1.82 & 0.041** & 0.020 & 0.002 \\
Llama DPO   & 0.833 & 0.019 & 1.90 & 0.102** & 0.044* & 0.010 \\
Llama SFT   & 0.130 & 0.009 & 1.13 & 0.156** & 0.035 & 0.024 \\
\bottomrule
\end{tabular}
\end{minipage}%
\hfill
\begin{minipage}[c]{0.44\columnwidth}
\caption{SFT vs.\ DPO fine-tuning comparison. Entr.\ = mean per-question Shannon entropy; Var = mean per-question variance; Eff.Cat = effective number of response categories ($2^{\text{entropy}}$); Mantel $r$ and Elem $r$ = correlation structure metrics vs.\ human; RV = RV coefficient. **$p<0.01$, *$p<0.05$.}
\label{tab:sft_vs_dpo}
\end{minipage}
\end{table}

SFT exhibits severe mode collapse: \qty{84.0}{\%} of all responses default to ``Accept,'' 78 of 100 questions have $>$\qty{90}{\%} identical answers, and 55 questions have zero variance. Mean entropy drops to 0.130 (\qty{10}{\%} of human), with an effective 1.13 response categories---effectively a single constant output. By contrast, DPO maintains 1.90 effective categories and preserves meaningful within-question variation.

While SFT achieves nominally higher element-wise Pearson $r$ (0.349 vs.\ 0.257 for DPO on the full matrix), this is inflated by mode collapse: when the model always predicts ``Accept'' and the human mode is also ``Accept,'' spurious correlation arises. Restricting evaluation to questions where the model exhibits non-trivial variance (var $> 0.01$) reveals only 13 qualifying questions for SFT versus 60 for DPO, with SFT's conditioned $r$ dropping to 0.090 (DPO: 0.248).

Table~\ref{tab:sft_collapse} quantifies the mode collapse across models: Llama-3.1-8B Base already exhibits a modest bias toward ``Accept'' (77.3\% modal, 33/100 near-uniform questions), which SFT amplifies into a near-constant output (97.0\% modal, 55/100 zero-variance, a 5.8$\times$ compression of human variance), whereas DPO preserves the base model's variance profile far more faithfully (74.7\% modal, 2.9$\times$ compression).

\begin{table}[t]
\begin{minipage}[c]{0.52\columnwidth}
\centering
\footnotesize
\setlength{\tabcolsep}{3pt}
\begin{tabular}{@{}lcccc@{}}
\toprule
Model & Mode\% & Q $>$\qty{90}{\%} & Zero-var & Col.${}\times$ \\
\midrule  
Human      & \qty{54.7}{\%} &  0 &  0 & $1.0\times$ \\
Llama Base & \qty{77.3}{\%} & 33 &  5 & $2.0\times$ \\
Llama DPO  & \qty{74.7}{\%} & 26 &  9 & $2.9\times$ \\
Llama SFT  & \qty{97.0}{\%} & 78 & 55 & $5.8\times$ \\
\bottomrule
\end{tabular}
\end{minipage}%
\hfill
\begin{minipage}[c]{0.44\columnwidth}
\caption{Mode collapse diagnosis. Mode\% = modal fraction; Q $>$\qty{90}{\%} = near-uniform questions; Zero-var = zero-variance questions; Col.${}\times$ = human-to-model variance ratio.}
\label{tab:sft_collapse}
\end{minipage}
\end{table}

These results confirm that DPO is more suitable than SFT for this domain. SFT's standard language modeling objective encourages imitating the majority label, producing near-uniform outputs that fail to capture philosophical disagreement. DPO's contrastive objective---which explicitly learns to \emph{distinguish} between preferred and non-preferred positions---provides a stronger learning signal per training example, preserving the ability to produce differentiated responses even with limited training data. This finding aligns with the observation of \citet{kirk2024understanding} that alignment methods reduce output diversity, but demonstrates that the severity varies markedly across training objectives.

\end{document}